# Factor of 1000 suppression of the depolarization rate in ultracold thulium collisions

I.A. Pyrkh[1,2], A.E. Rudnev[1,2], D.A. Kumpilov[1,2], I.S. Cojocaru[1,3], V.A. Khlebnikov[1], P.A. Aksentsev[1,4], A.M. Ibrahimov[1,2], K.O. Frolov[1,2], S.A. Kuzmin[1,2], A.K. Zykova[1], D.A. Pershin[1], V.V. Tsyganok[1], A.V. Akimov[1,3]

*[1]Russian Quantum Center, Bolshoy Boulevard 30, building 1, Skolkovo, 121205, Russia*

*[2]Moscow Institute of Physics and Technology, Institutskii pereulok 9, Dolgoprudny, Moscow Region 141701, Russia*

*[3]PN Lebedev Institute RAS, Leninsky Prospekt 53, Moscow, 119991, Russia*

*[4]Bauman Moscow State Technical University, 2-nd Baumanskaya, 5, Moscow, 105005, Russia*

*Email: a.akimov@rqc.ru*

Lanthanides are nowadays extensively used to investigate the properties of strongly correlated matter. Nevertheless, exploiting the Zeeman manifold of a lanthanide atom ground state is challenging due to the unavoidable presence of depolarization collisions. Here we demonstrate that in the case of the thulium atom, it is possible to suppress this depolarization by a factor of 1000 with a carefully tuned magnetic field thus opening the way for the efficient use of the Zeeman manifold in quantum simulations.

Since the achievement of the Bose-Einstein condensation of rubidium [1], sodium [2], and lithium [3] atoms, ultracold gases of bosons have attracted significant attention as a test-bed system for exploring the properties of strongly correlated matter. The character of interactions between particles affects the properties of a quantum gas. Lanthanides, due to an incomplete $f$ shell, have a dense spectrum of Feshbach resonances [4–14] causing the strong dependence of interatomic scattering on the magnetic field. Large magnetic moment of lanthanides introduces the dipole-dipole interaction in the system [15–19]. In particular, thulium atom with one hole in $f$-shell has $4\mu_B$ magnetic moment in the ground state as well as rich spectrum of Feshbach resonances [20], that is less dense compared to such lanthanides as erbium [21] and dysprosium [22]. These aspects along with a relatively simple level structure make thulium a convenient ultracold research tool.

The anisotropy and the long range of dipole-dipole interactions (DDI) modify the scattering properties and many-body behavior making it possible to observe, in particular, quantum droplets [23–26] and supersolids [27–29]. The large spin manifold of highly magnetic atoms makes it possible to exploit spin richness to explore phenomena like quantum magnetism [30] and exotic topological phases [31,32], to realize long-range-interacting spin-lattice models [19,33–35], and to study non-classical spin states within a highly controlled system [36,37]. Exploiting internal degrees of freedom allows to create artificial field [33], study dark lattices in Gauge theories [34], as well as feminizations of bosons, which can happen in specially arranged lattices or special interaction regimes [35] and potentially formation of magnetic phases previously mostly considered with fermions [36,37].

When two magnetic atoms collide, the projections $m_{F1}$ and $m_{F2}$ of their respective total angular momenta $\vec{F_1}$ and $\vec{F_2}$ may change after the collision. Collisions in which $m_F = m_{F1} + m_{F2}$ is decreased (relaxation collision) usually lead to the loss of the atomic pair from the trap, as the Zeeman energy transfers to the kinetic energy of atoms. Collisions in which $m_F$ is conserved (depolarization collision) leave atoms in the trap but lead to the depolarization of the initial state. We note, that this mechanism is not the only possible mechanism of depolarization. In case of atoms with zero orbital momentum in the ground state it has been demonstrated, that spin-exchange interaction may play significant role [38], as well as dynamics of evaporation cooling [39] in case of fermions. Nevertheless, the significant magnetic moment in the ground state makes the dipolar mechanisms to dominate in the depolarization.

Preparing the atoms in the stretched state $m_F = 2F$ eliminates the depolarization collisions. The dipolar relaxation from the stretched state can be reduced by confining atoms in thin layers [40], loading into optical lattices or applying RF fields [41]. External magnetic field $B$ also modifies the relaxation rate in the stretched state. The Born approximation of the DDI potential **Ошибка! Источник ссылки не найден.** gives $\propto \sqrt{B}$ dependency for the relaxation cross-sections in the case of two identical bosons and $\propto 1/\sqrt{B}$ in the case of identical fermions, both were experimentally demonstrated for Dy isotopes [42] Noticeably, the relaxation cross-section of bosonic Cr exhibits a dip in the region of small magnetic field, that was described by including interatomic potential and applying Fermis golden rule with the numerically calculated wavefunctions [41].

When atoms aren't initially prepared in one of the stretched states $m_F = \pm 2F$, depolarization collisions emerge. Although they conserve both Zeeman energy and kinetic energy of atoms, they complicate the description of the dipolar collisions. The depolarization can be mitigated in 3D optical lattices [43] or by applying different AC Stark shifts to the Zeeman sublevels [44]. In the latter case the Dy bosons in the state $|J_1 = 8, m_J = -7\rangle$ exhibit the dip in the relaxation cross-section similar to the Cr mentioned above.

Here we demonstrate the existence of a magnetic field-dependent resonance, suppressing the depolarization rate in thulium by more than 1000 times. This type of suppression to our knowledge never been neither predicted nor observed before. The well-established radio-frequency protocol for spin state initialization [45] was used to manipulate the spin mixture and probe the population of different sublevels. The evolution of populations made it possible to determine the collision rates that strongly depend on the magnetic field.

The experiment starts with the production of an ultracold gas of thulium atoms detailed in previous works [18,20,46–49]. The specific microwave part of the setup is exact copy of the one presented at [45]. Thulium has only one stable isotope $^{169}$Tm with $LS$ bond in the ground state $4f^{13}\left({}^2F^o\right)6s^2$ ($L=3,\ S=1/2,\ J=7/2$) which has two hyperfine components $F=4$ (lower) and $F=3$ (upper) due to the non-zero nuclear spin $I=1/2$. The Zeeman slower and 2D optical molasses operating at the strong transition $4f^{13}\left({}^2F^0\right)6s^2 \rightarrow 4f^{12}\left({}^3H_5\right)5d_{3/2}6s^2$ with a wavelength of 410.6 nm and a natural width of $\Gamma = 2\pi\gamma = 2\pi \cdot 10.5\,\text{MHz}$ provided the precooling of atoms. Then the atoms were loaded into the magneto-optical trap operating at the weaker transition $4f^{13}\left({}^2F^o\right)6s^2 \rightarrow 4f^{12}\left({}^3H_6\right)5d_{5/2}6s^2$ with a wavelength of 530.7 nm and a natural width of $\Gamma = 2\pi\gamma = 2\pi \cdot 345.5\,\text{kHz}$. The reduction of magneto-optical trap light intensity provided the polarization of atoms at the lowest magnetic sublevel $|F=4; m_F = -4\rangle$ of the ground state. After cooling down to 22.5 ± 2.5 μK, the atoms were loaded into the optical dipole trap formed by a linearly polarized laser beam waist of 40 μm and a wavelength of 1064 nm. The subsequent evaporation finally produced about $2.0 \cdot 10^6$ atoms with 2.5 μK in the trap with $\left(\omega_x, \omega_y, \omega_z\right) = 2\pi\left(326, 326, 2\right)$ Hz frequencies.

In the presence of an external magnetic field, both hyperfine components split into the Zeeman manifolds (see Figure 1a) separated by frequencies $g_F B \mu_B / h$, where $g_F$ is the Lande g-factor of $F$ level, $B$ is the absolute value of the magnetic field, $h$ is the Planck constant, and $\mu_B$ is the Bohr magneton. Since $g_F$ takes different values for the hyperfine components ( $g_{F=4} = 0.999$ and $g_{F=3} = 1.284$ ), a non-zero magnetic field splits their Zeeman manifolds differently thus allowing frequency selective addressing of every specific $|F=4\rangle \to |F=3\rangle$ transition with a MW field using an antenna designed previously [45]. Figure 1a presents a sequence of the two $\pi$-pulses populating the $|4,-3\rangle$ spin state. Similarly, a sequence of $\pi$-pulses pairs with adjusted frequencies and duration can populate any specific Zeeman sublevel of the ground state in the ladder-like manner.

In the lowest $|4,-4\rangle$ spin state, the depolarization and relaxation processes are suppressed. Contrarily, in the $|4,-3\rangle$ spin state, both processes are allowed. The corresponding spin dynamics was investigated exploiting the $|4,-3\rangle \to |3,-3\rangle$ and $|3,-3\rangle \to |4,-3\rangle$ $\pi$-pulses (35 and 50 μs duration, respectively). Figure 1b presents the three pulse sequences which provide the number of atoms $N_{tot}$, $\overline{N}_3$ and $\overline{N}_4$ after the absorption imaging. The two subsequent $|4,-3\rangle \to |3,-3\rangle$ and $|3,-3\rangle \to |4,-3\rangle$ $\pi$-pulses transferred all the atoms to the $|4,-3\rangle$ spin state. The probe beam was tuned to the wide 410.6 nm transition so the absorption imaging would contain all the atoms in the $|F=4\rangle$ Zeeman manifold. Therefore, the $N_{tot}$ is the total population of all sublevels of $|F=4\rangle$, the $\overline{N}_3$ is the number of atoms in all the sublevels except $|4,-3\rangle$, and the $\overline{N}_4$ is the number of atoms in all the sublevels except $|4,-4\rangle$ after the $\tau$ time period.

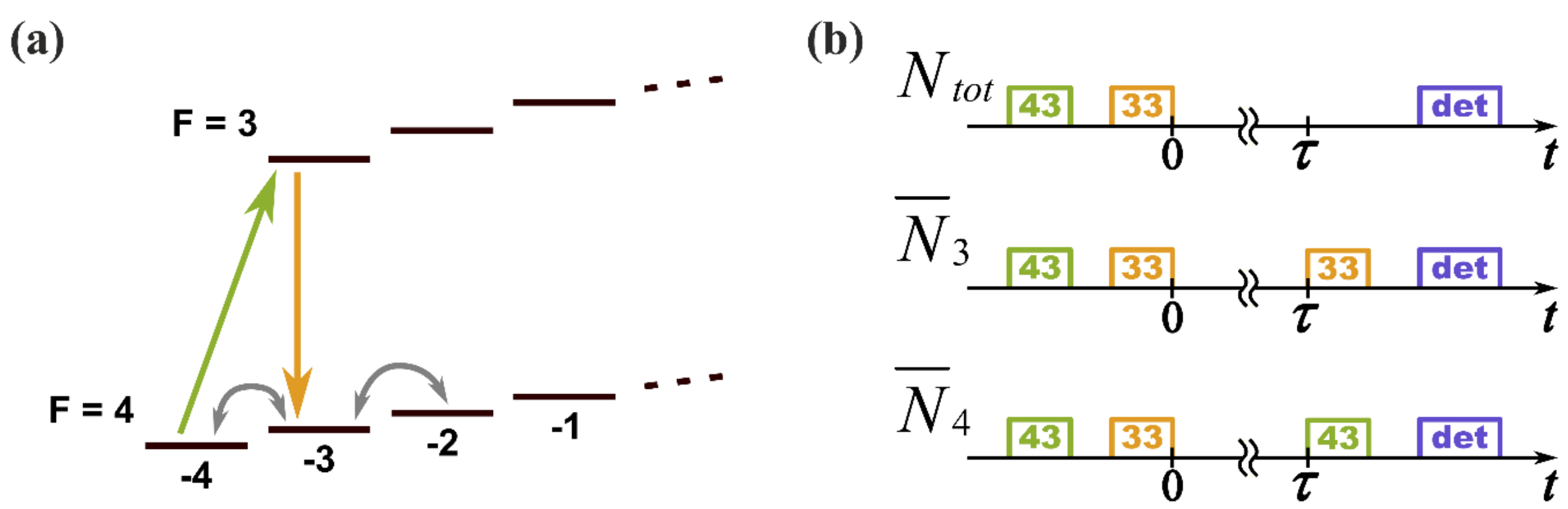

Figure 1. a) The scheme of experimental pulses. The colored arrows present $\pi$-pulses, and the gray curved arrows represent spin-exchange and relaxation channels. b) The scheme of the microwave $\pi$-pulses driving the selected transitions to obtain the $N_{tot}$, $\overline{N}_3$ and $\overline{N}_4$.

Figure 2 presents the experimentally measured $N_{tot}(\tau)$, $\overline{N}_3(\tau)$ and $\overline{N}_4(\tau)$ dependancies, (a) and (b) corresponding to $B = 2.98\,G$ and $B = 0.565\,G$ respectively. For the smallest $\tau$, the $\overline{N}_3$ is predictably negligible, because all the atoms were in the $|4,-3\rangle$ state and were removed from the $|F=4\rangle$ manifold. Correspondingly, the $\overline{N}_4$ was similar to the $N_{tot}$. However, for larger $\tau$, the $\overline{N}_3$ increases revealing a gradual appearance of atoms not in the $|4,-3\rangle$ state. Correspondingly, the $\overline{N}_4$ becomes smaller than $N_{tot}$ demonstrating the gradual appearance of atoms particularly in the $|4,-4\rangle$ state. This behavior indicates the presence of spin dynamics processes.

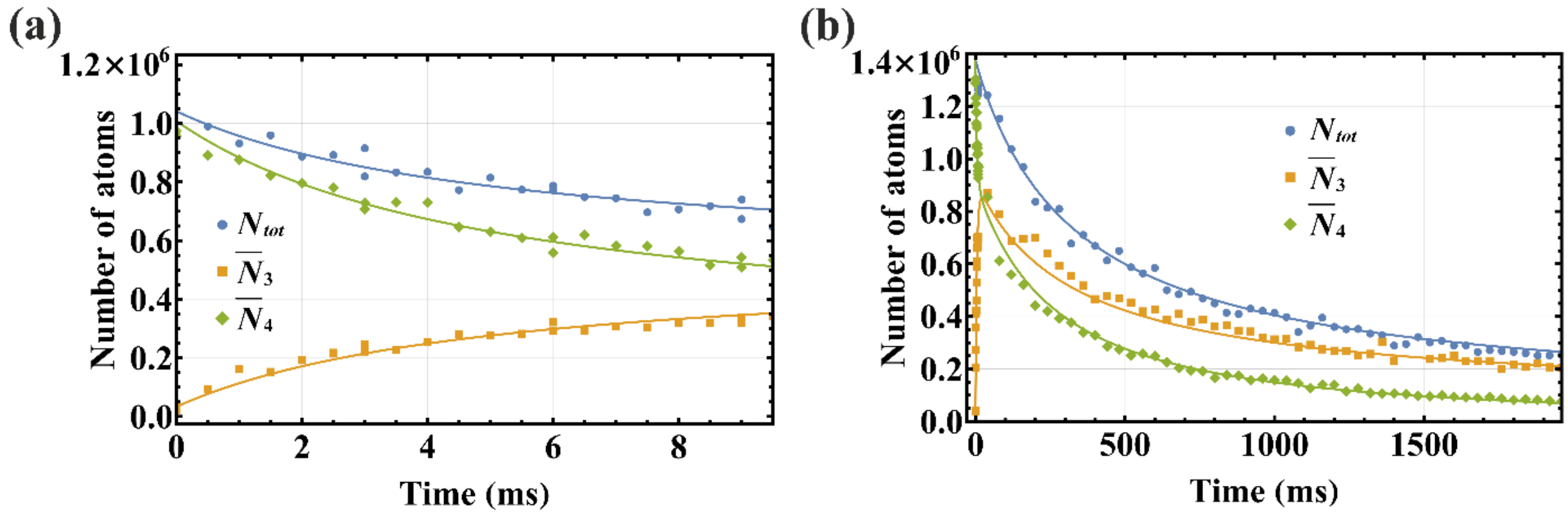

Figure 2. The number of detected atoms versus time: (a) in the magnetic field $B = 2.98\,G$ at short time and (b) in $B = 0.565\,G$. Every experimental value was obtained from a single absorption image. Solid lines in (a) correspond to the fit by system (2) accounting for the relations (3) providing the $N_{tot}$, $\overline{N}_3$ and $\overline{N}_4$ values. Solid lines in (b) correspond to the fit by the system with next order collisions (see Supplementary Materials)**Ошибка! Источник ссылки не найден.**.

Considering only the two-body collisions, there are depolarization processes like $|4,-3\rangle \otimes |4,-3\rangle \rightarrow \frac{1}{\sqrt{2}}\left(|4,-4\rangle \otimes |4,-2\rangle + |4,-2\rangle \otimes |4,-4\rangle\right)$ and relaxation processes leading to losses from the trap. If we consider the dynamics only at short times (while populations of the all states but $|4,-3\rangle$ are negligeble), then the number of atoms in the $|4,-4\rangle$ and $|4,-2\rangle$ states is relatively small and only collisions of atoms in $|4,-3\rangle$ state are important. In this case, we can describe dynamics by the system

$$\begin{cases} \dot{N}_3 = -\alpha(\beta_{loss} + \beta_{depol})N_3^2 \\ \dot{N}_2 = \dot{N}_4 = \alpha \dfrac{\beta_{depol}}{2} N_3^2, \end{cases} \tag{1}$$

where $\alpha = \dfrac{\bar{\omega}^3}{(2\pi)^3}\left(\dfrac{\pi m}{kT}\right)^{\frac{3}{2}}$, $\bar{\omega} = \left(\omega_x\omega_y\omega_z\right)^{\frac{1}{3}}$, $T$ is the atomic cloud temperature and $m$ is the mass of atom. The $N_4$, $N_3$, and $N_2$ are the populations in the states $|4,-4\rangle$, $|4,-3\rangle$, and $|4,-2\rangle$, respectively, $\beta_{depol}$ is a depolarization spin-exchange rate, $\beta_{loss}$ — spin-relaxation rate leading to losses from the trap. The experimental data correspond to

$$\begin{aligned} N_{tot} &= a_4N_4 + a_3N_3 + a_2N_2 \\ \overline{N}3 &= a_4N_4 + a_2N_2 \\ \overline{N}4 &= a_3N_3 + a_2N_2 \end{aligned} \tag{2}$$

where $a_4$, $a_3$ and $a_2$ are the detection efficiencies for $N_4$, $N_3$, and $N_2$, respectively (see Supplementary Materials). Figure 2a demonstrates the numerical solution to the system (1) fitted the experimental decay data calculated according to (2) with parameters $N_{03}$, $N_{04}$, $\beta_{depol}$ and $\beta_{loss}$ with initial conditions $N_3(0) = N_{03}$, $N_4(0) = N_{04}$, $N_2(0) = 0$.

At the long times system (1) can be appended with next order processes (see Supplementary Materials) While large number of additional parameters does not allow to extract all these additional parameters, one could see contribution of additional channels by fitting data at long times (see Figure 2b). Yet due to the complication of this model, only equations (1) were used for further analysis.

The rates obtained from fitting the data by system (1) are presented in Figure 3. Both the depolarization and loss rates demonstrate a resonance-like and non-monotonic behavior, particularly revealing the 0.90 G value of the magnetic field at which the $\beta_{depol}$ rate is more than 1000 times less than the $2\cdot10^{-11}$ baseline (Figure 3). The depolarization suppression resonance is also reproduced in the loss rate $\beta_{loss}$ (Figure 3b) by a factor of 50 at the same magnetic field. Overall, the loss rate curve closely resembles the depolarization curve in shape, but appears compressed along the vertical axis.

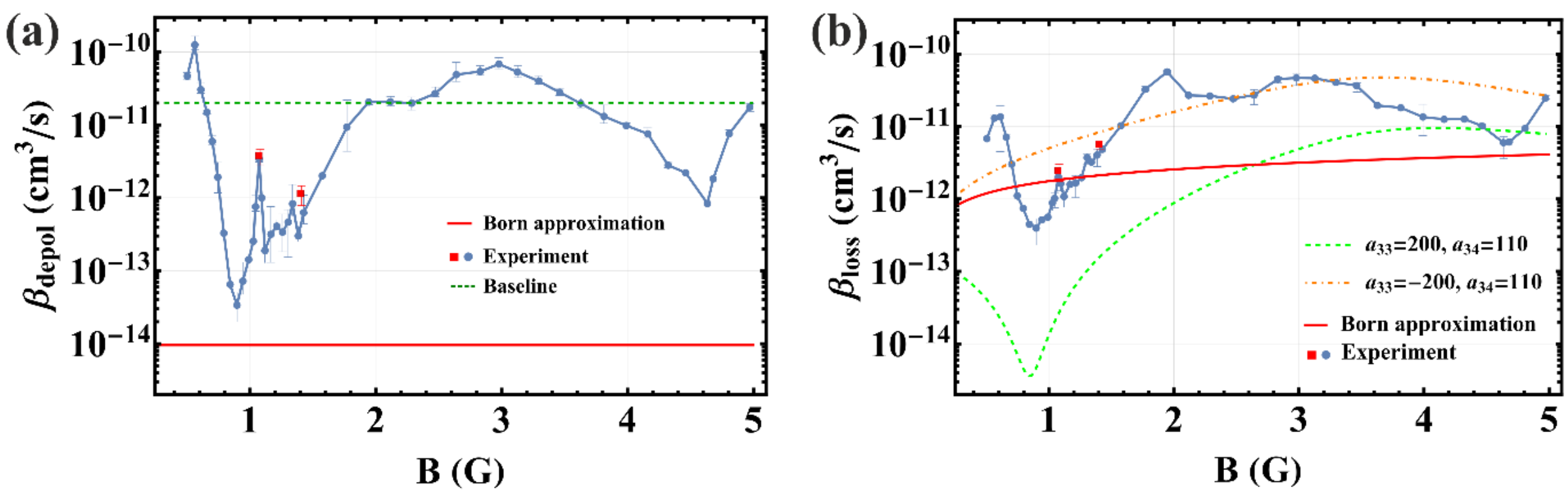

Figure 3. a) The dependence of $\beta_{depol}$ on the magnetic field. The green dashed line illustrates the baseline $2 \cdot 10^{-11}$. The red line represents the values obtained from the Born approximation. The red squares and blue dots are experimental values with and without additional $\overline{N}_2$ data, correspondingly. b) The dependence of $\beta_{loss}$ on the magnetic field. The red line is the Born approximation values. The green dashed line and orange dot-dashed line are theoretical predictions based on model from [41] (see Supplementary Material). The green dashed curve corresponds to the values of $a_{33} = 200,\ a_{34} = 110,$ (given in Bohr radii) orange dot-dashed curve corresponds to $a_{33} = -200,\ a_{34} = 110$. Red squares and blue dots have the same meaning as in (a).

To validate the model (1), additional experiments were performed with a $|4,-2\rangle \to |3,-1\rangle$ $\pi$ -pulse to obtain the $\overline{N}_2$ data (see Supplementary Materials). Loss rate values $\beta_{loss}$ and $\beta_{depol}$ were calculated with and without regard to the additional data. The obtained values for two magnetic fields are shown in Figure 3 as red squares and coincide with the initial experimental values within error bars.

To check the validity of the two-body model, a series of experiments similar to the one in Figure 2 with different initial atomic numbers were conducted. Given the constant ODT volume and the same starting temperature, we can explore the dependence of $\beta_{loss}$ and $\beta_{depol}$ on the initial number of atoms. Figure 4 shows that $\beta_{loss}$ and $\beta_{depol}$ are close to constant values, $\beta_{loss} = (2.6 \pm 2.3) \times 10^{-12}$ and $\beta_{depol} = (1.5 \pm 0.4) \times 10^{-11}$, and thus the two-body collision model does describe the experiment in a self-consistent way.

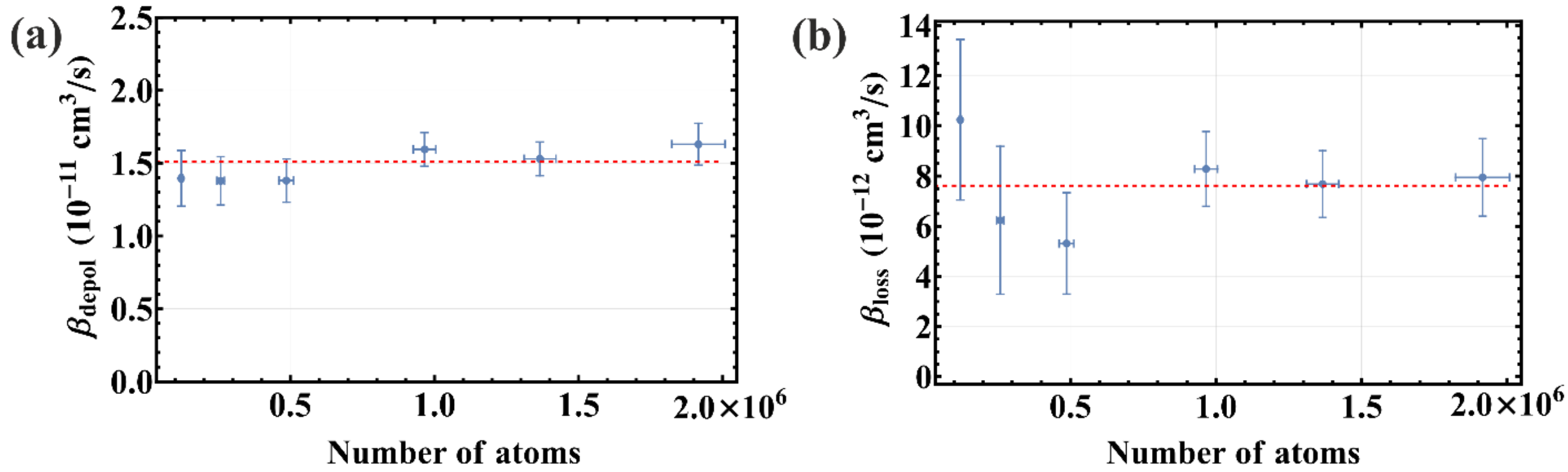

Figure 4. The depolarization rate $\beta_{depol}$ (a) and loss rate $\beta_{loss}$ (b) as functions of the initial atomic numbers in the magnetic field $B = 0.66\,G$.

Applied to the relaxation cross-section the Born approximation (see Supplementary Material) gives values lower than most experimental ones (the red line in Figure 3b). Including the interatomic potentials by applying the Fermis golden rule requires to adjust the scattering lengths $a_{33}$ and $a_{34}$, the value $a_{44} = 144a_B$ was measured experimentally and fixed. The $a_{33} = 200a_B$, $a_{34} = 110a_B$ (green dashed curve) qualitatevely gives the dip in the magnetic field 0.90 G, but fails to reproduce large magnituteds of $\beta_{loss}$. Larger $\beta_{loss}$ was achieved for $a_{33} = -200a_B$, $a_{34} = 110a_B$ (orange dot-dashed line), but there are no values $a_{33}$, $a_{34}$ for which theoretical curve fit all the experimental data points.

For depolarization collisions the Born approximation gives cross-section independent on the magnetic field (the red line in Figure 3a). In contrast to this, in the experiment we see that depolarization has clear non-monotonic behavior with several resonances and up to 4 orders of magnitude higher values. Thus, we conclude that perturbational approaches are insufficient to describe observed collisional rates. In fact, it is immediately evident from strong correlation between depolarization and relaxation rates, that one needs to account for coupling between different states. This requires carrying out coupled-channels calculations. For the present dipolar interactions they were done for hydrogen [50] and in general can be carried out for alkali metals using packages as MOLSCAT [51,52]. Such calculations require intricate knowledge of the interatomic potential curves making them hard to realise for lantanides, because they have complex angular momentum structure leading to large amount of Born-Oppenheimer potentials [53]. This leads to the dense spectrum of Feshbach resonances which occur when energy of the energetically closed channel coincides with the energy of the atomic pair in the entrance channel and subsequent chaotic dependence of the scattering on interaction parameters [20]. The vicinity of Feshbach resonances can both enhance [54] and suppress [55] two-body inelastic collision rates, that probably explains the observed non-monotonic behavior.

In the magnetic field of 0.90 G the lifetime of the $|4,-4\rangle$ state is around 6.8 s. At the low values of $\beta_{loss}$ and $\beta_{depol}$ system (1) is sufficient to describe spin dynamics at all times, so the lifetime of the $|4,-3\rangle$ state can be defined as $\tau = \frac{e-1}{\alpha(\beta_{loss}+\beta_{depol})N_3(0)}$ and is around 2.3 s that is only 3 times shorter compared to $|4,-4\rangle$ state.

Thus, the developed experimental procedure can be exploited to populate specific Zeeman sublevels and investigate spin dynamics. Particularly, in ultracold thulium at the magnetic field of 0.9 G, both the depolarization rate and the loss rate for the $|4,-3\rangle$ state are significantly suppressed by a factor of 1000 and 50, respectively, providing considerable lifetime of this state. Such features pave the way for utilizing the Zeeman manifold of the thulium atom to explore the properties of strongly correlated matter.

## ACKNOWLEDGMENTS

This work of was supported by Rosatom in the framework of the Roadmap for Quantum computing (Contract No. 868-1.3-15/15-2021 dated October 5).